\documentclass[a4paper,fleqn,usenatbib]{mnras}

\usepackage{graphics,epsf}
\usepackage{amsmath}                
\usepackage{amsfonts}               
\usepackage{amssymb}                
\usepackage{epsfig}                 
 
\title[Dust-driven common-envelope ejection]{Efficient common-envelope ejection through dust-driven winds}

\author[H. Glanz \& H. B. Perets]{
Hila Glanz and Hagai B. Perets 
\\
Technion - Israel Institute of Technology, Haifa, 3200002, Israel}

\date{Accepted XXX. Received YYY; in original form ZZZ}

\pubyear{2018}

\begin{document}
\label{firstpage}
\pagerange{\pageref{firstpage}--\pageref{lastpage}}
\maketitle

\begin{abstract}
Common-envelope evolution (CEE) is the short-lived phase in the life
of an interacting binary-system during which two stars orbit inside
a single shared envelope. Such evolution is thought to lead to the
inspiral of the binary, the ejection of the extended envelope and
the formation of a remnant short-period binary. However, detailed
hydrodynamical models of CEE encounter major difficulties. They show
that following the inspiral most of the envelope is not ejected; though
it expands to larger separations, it remains bound to the
binary. Here we propose that dust-driven winds can be produced following
the CEE. These can evaporate the envelope following similar processes
operating in the ejection of the envelopes of AGB stars. Pulsations in an AGB-star drives the expansion of its envelope, allowing the material to cool down to
low temperatures thus enabling dust condensation. Radiation
pressure on the dust accelerates it, and through its coupling to the
gas it drives winds which eventually completely erode the envelope.
We show that the inspiral phase in CE-binaries can effectively replace
the role of stellar pulsation and drive the CE expansion to scales
comparable with those of AGB stars, and
give rise to efficient mass-loss through dust-driven winds. 
\end{abstract}




\section{Introduction}

Post common-envelope (CE) binaries are thought to give rise to many
types of interacting binary systems. These include the progenitors
of transient explosive events, such as Type Ia supernovae, short gamma
ray bursts, X-ray binaries and double WD/NS/BH systems Inspiraling
through GW emission. However the evolved components of such systems
must have once been orders-of-magnitude larger than would fit within
the size of the observed present-day systems. The observations of
such systems therefore require some migration process to drive these
binaries to their current compact configuration.

Common-Envelope Evolution (CEE) is the short-lived phase in the life
of an interacting binary star during which two stars orbit inside
a single, shared envelope. The interaction of the stellar components
with the envelope through gravitational torques leads to their inspiral
into a short period orbit, and is thought to also lead to the ejection
of the common envelope. CEE is therefore currently accepted as the
evolutionary process allowing the formation of compact short-period
binaries (see \citealt{Pod01,Izz+12,Iva+13,Sok17} for reviews). However,
detailed hydrodynamical models of CEE encounter major difficulties;
they show that following the in-spiral phase most of the envelope
is \emph{not} ejected, but only expands to larger separations where
it still remains bound to the binary (e.g. \citealt{Ric+12,Iva+13,Iva+15,Kur+16,Ohl+16,Iac+17}).

It was suggested \citep[and references therein]{Nan+15,Cla+17} that recombination  
energy stored in ionized hydrogen and helium may provide an additional 
energy source driving envelope ejection. Such a scenario follows similar 
models suggested in the past for the ejection of the envelope of single 
AGB stars (e.g. \citealt{Rox67,Pac+68,Han+94}). In this context, the fraction of the recombination energy lost to radiation is still debated and is actively studied \citep{Sok+03, 2011ASPC..447...91I, Cla+17, 2017MNRAS.472.4361S}. It is also still not clear whether this channel can eject the CE in all cases \cite{Cla+17}, and whether it can explain wide post-CE orbits \cite{Iva+15}. 
Irrespectively of this channel, one can follow 
a different path in searching for mechanisms that eject the envelope
in single asymptotic giant-branch (AGB) stars, and extend them to
CEE. In particular, we propose to follow the current paradigm for
the ejection of the envelope of an AGB star, namely dust-driven winds
\citep[see][for a review]{Lam+99}, and apply it for CEE scenarios.

Mass loss in AGB stars is thought to proceed through slow stellar
winds. Stellar pulsations levitate material outwards sufficiently
far as to allow for its cooling to low temperatures at which dust
condensation becomes possible. The dust grains are accelerated away
due to radiation pressure from the star, and through their collisional
coupling to the gas, they accelerate the gaseous envelope in which
they are embedded, and eventually evaporate the entire envelope (e.g.
\citealt{Win+00,Hoff+16}).

We propose that such dust-assisted radiative envelope evaporation
mechanism \textbf{\emph{can be similarly effective}} in evaporating
CEs of both red giants (RG) and AGB binaries. The possible importance
of dust for CEE was noted in earlier studies (\citealt{Sok92,Sok04}),
but had not been studied in depth. In CE hydrodynamical simulations
the primary giant star expands significantly, but loses only a fraction
($10-25\%$) of its mass, while most of the mass remains bound to
the star \citep{Ric+12,Iva+13,Iva+15,Kur+16,Ohl+16,Iac+17}. The size
scale of the bloated bound-envelope is comparable to that of an AGB
envelope, and contains comparable mass. The in-spiral of the companion
naturally extends the envelope (and may also form pulsations in the
CE; \citealt{Sok92}), replacing an important part of the role of
pulsations considered in single AGB stars (e.g. \citealt{Bow88,Sok92}).
Indeed as can be seen in CE hydrodynamical simulations \citep{Ric+12,Iva+13,Iva+15,Kur+16,Ohl+16,Iac+17},
the densities in the CE at this stage are comparable to those of an
AGB star. As the envelope expands and cools down, the in-spiral induced
pulsations give rise to shocked, higher density regions in the CE.
In other words, the CE evolution gives rise to conditions very similar
to those existing in AGB envelope, and in particular, conditions that
allow for dust formation.

In the dust-assisted CEE suggested scenario the inspiral of the companion
in a CE induces the expansion of the red-giant envelope, which then
cools and forms dust; the radiation pressure from the remnant stellar
core can accelerate the dust through radiation pressure and the collisional
coupling of the dust to the gas can then give rise to the ejection
of the entire CE.We emphasize that the inspiral phase is critical for this process since it leads to the expansion and cooling of the atmosphere, which can only then give rise to dust formation. We note that this does not need to be an alternative
to other scenarios, such as the recombination mediated CEE, but might
even assist other processes. For example, if dust forms at some distance
above the recombination region, the much larger opacity it produces
could be somewhat more efficient in trapping of any available recombination
energy and thereby overcome the potential challenges raised in this
context\footnote{In fact, this issue of dust-assisted in the context of recombination
energy was ignored for the ejection of AGB envelopes, but it might
be important in that context too; this would be explored elsewhere.}. Any recombination radiation may also assist in producing radiation
pressure on the dust. In the following we discuss the dust-driven
CEE scenario in detail.

We begin by describing the mechanism for dust driven winds, as developed
in the context of AGB stars (section \ref{sec:dust-driven-winds}.
We then discuss the relevant properties of CE binaries and the condensation
radius above which dust may form in a CE. Finally we discuss dust-driven
winds ejection of a CE, and provide a worked-out example for a typical
case (section \ref{sec:CE}), and then summarize.

\section{Dust Driven Winds}

\label{sec:dust-driven-winds}

The current paradigm for the mass loss in in AGB stars and luminous
red super-giants is through dust-driven winds \citep[see][for a review]{Lam+99}.
As these stars evolve they experience significant pulsations that
levitate material far beyond the stellar surface, where it can cool
and condense into dust grains. The grains absorb and scatter stellar
photons from the star, and are accelerated outwards. Collisional coupling
of the dust grains and the surrounding gas result in an effective
radiation pressure, mediated by dust, on the stellar envelope. Given
the large opacity of the dust grains, the evolved star effectively
becomes super-Eddington, and the gaseous envelope, accelerated beyond
the escape velocity of the star, becomes unbound and is ejected through
a slow wind (with velocities comparable to the escape velocity from
the star).

Following \citet{Lam+99} one can define, $\Gamma_{d}$ the ratio
of radiative acceleration to gravity. In case of radiation forces
on dust, this ratio is $\Gamma_{d}=k_{rp}L_{\star}/4\pi GM_{\star}$,
where $k_{rp}$ is the radiation pressure mean opacity, and ${\rm L}_{\star}$
and ${\rm M}_{\star}$ are the luminosity and mass of the star, respectively.
Winds can occur if $\Gamma_{d}$ increases outwards and becomes greater
than unity, beyond some radius. The limiting luminosity for such dust-driven
winds can then be derived by taking $\Gamma_{d}$ to unity to obtain
\begin{equation}
L_{\star}>\frac{4\pi cGM_{\star}}{k_{rp}}\simeq400L_{\odot}\frac{M_{\star}/M_{\odot}}{k_{rp}/30{\rm cm^{2}g^{-1}}}\label{eq:1}
\end{equation}
for mean opacity $k_{rp}$, derived from detailed calculations over
many types of grains. The survival of a test grain can then be determined
by deriving its radiative equilibrium temperature. If the dust temperature
exceeds the condensation temperature the grain will sublimate rather
than grow. In addition, the density should be sufficiently high as
to allow for accumulation of solids contribution to the dust growth.
Effectively these conditions require low-temperature, but high density
environments, which can be attained in the envelope of pulsating AGB
star. As we discuss in the following section, similar conditions also
exist in CE binaries.

The distance from the star at which dust-grain temperature is below
the condensation temperature is independent of the density of the
ambient gas. As explained in details in \citet{Lam+99}, the low equilibrium
temperature is achieved when the condensation radius, represented
by $R_{C}$ has the following relation with the condensation temperature,
$T_{C}$:

\begin{equation}
W\left(R_{C}\right)=\left(\frac{T_{C}}{T_{\star}}\right)^{p+4},\label{eq:2}
\end{equation}
where the opacity is assumed to follow a power-law dependence $\kappa\propto\lambda^{-p}$
(where the specific relation depends on the grains properties such
as composition and size distribution), and $R_{\star},\,T_{\star}$
are the stellar radius and temperature, respectively. $W\left(r\right)$
is the 'geometrical dilution factor' and depends on the optical depth
at the radius. In most cases of AGBs, the optical depth of the gas
in the region of the condensation radius is much lower than unity
(the intensity incident on the dust grains is direct star light) and
thus $W\left(R_{C}\right)\approx\left({\rm R_{\star}/3{\rm R_{C}}}\right)^{2}$,
and as a consequence: 
\begin{equation}
R_{C}\approx\left(\frac{R_{\star}}{2}\right)\left(\frac{T_{C}}{T_{\star}}\right)^{-\frac{4+p}{2}}.
\end{equation}
The condensation radius is typically a few $\sim2-3$ times the stellar
radius of the evolved star \citep{Hoff+08}. While the temperature
and radiation field determine whether grains can form and at a given
distance from the star, it is the density at the dust formation radius
that determines the mass loss rate of the wind. The density determines
whether there is sufficient momentum coupling between grains and the
gas. Efficient mass-loss can maintained as long the atmospheric density
scale height can be significantly increased relative to that of gas
pressure supported hydrostatic atmospheres. A critical point is attained
when the gas velocity exceeds the sound speed, and enhanced density
can be developed through shocks. This may occur just before $\Gamma_{d}$
becomes larger than unity. Once the grains form, they are accelerated
through radiation pressure, and through their strong coupling to the
gas, they accelerate the gas up-to supersonic velocities. The gas
density then drops rapidly with the radius and further dust growth
is inhibited. This feedback provides a mechanism for conservation
of the grain sizes, and their effectiveness in inducing mass-loss.
In case of a thick shell, $\Gamma_{d}$ will be larger than 1 throughout
the region, where the temperature decreases between the condensation
temperature and $T_{surf}$ where $\Gamma_{d}=1$.

\section{Dust-assisted common-envelope ejection }

\label{sec:CE}

During the CE phase of a RG-binary, the companion spirals inside the
envelope due to gravitational torques, transferring angular momentum
and energy to the envelope of the evolved star, leading to a substantial
expansion of the envelope. Later, the spiral-in slows down, leaving
behind a bloated star containing the remnant core of the RG, which
radiates on the now extended gaseous envelope. This process transforms
the conditions in the RG envelope to be far more similar to those
existing in pulsating AGB stars, and thereby allows for similar production
of dust-driven winds.

Although the original temperature profile of the RG is too high for dust formation, the CE expands significantly during the CE phase, and thus far from the core, it is sufficiently cool but still dense enough as to allow for dust condensation. Indeed, previous theoretical
studies suggested the formation of dust in the ejecta of CEs \citet{Lu+13}.These, however, explored dust formation {\emph after} the ejection of the envelope in the unbound material, while our focus is on the initially bound material following its expansion due to the inspiral. The conditions in the latter case differ from those in the ejected material.
We aso note that direct observations show the existence of a significant amount
of dust shrouding/obscuring post CE/merger objects \citep{Kam+10,Kam+11,Tyl+11,Bar+14,Tyl+15}.
The radiation from the inner parts of the star applies radiation pressure
on the dust, accelerating it above the sound speed of the gas, and
the collisional coupling of the dust to the gas can eventually result
in the ejection of the entire outer layer of the envelope. Without
the existence of the dust - the extended envelope would eventually
cool down, release the energy as radiation and then fall back. Let
us now consider this in more detail.

\subsection{The dust condensation radius}

\label{subsec:cond-radius}

The luminosity of the RG stellar core is conserved during the CE phase,
and its source is the nuclear reactions in the star's core, which
do not change during the process. The luminosity of the RG companion
(a MS or a compact object) can typically be neglected compared with
that of the RG core; hence we make a simplified assumption that the
total luminosity in some given layer of the extended envelope is comparable
to the core luminosity. If we consider the inner part of the star
below the condensation radius as a black-body (the optical depth is
large) we can estimate the effective temperature at the condensation
radius, by using the initial values (prior to the CE) of the luminosity
($L_{\star}^{i}$), the radius ($R_{\star}^{i}$) and the effective
temperature ($T_{\star}^{i}$) of the RG star at this radius: 
\begin{equation}
\frac{L_{\star}^{i}}{L\left(r\right)}=\frac{4\pi\sigma R_{\star}^{i2}T_{\star}^{i4}}{4\pi\sigma r^{2}T\left(r\right)^{4}}\rightarrow T\left(r\right)=\sqrt{\frac{R_{\star}^{i}}{r}}T_{\star}^{i}\label{eq:4}
\end{equation}
Together with Eq. (2) (with $p\approx1$) and the optically thick
approximation ($W\left(R_{c}\right)\approx1 / 2$), the condensation
radius can be calculated as follows:
\begin{equation}
R_{C}=0.5^{2 / 5}\left(\frac{T_{C}}{T_{\star}^{i}}\right)^{-2}R_{\star}^{i}\label{eq:5}
\end{equation}

Usually the condensation radius of an AGB is located outside the stellar
surface, thus the pulsations play an important role in the gas ejection
process. The pulsations determine the rate at which gas from the star
reaches the condensation radius before being pushed away, and thereby
determine the dust creation rate. In the post-CE case the inspiral,
and possibly recombination effects, may also give rise to pulsational-like
behavior (e.g. \citealt{Cla+17}, but irrespective,
the resulting extended envelope is already far from the RG core, and
it is sufficiently cool and dense as to allow for dust condensation.
Indeed, the initial effective temperature, $T_{\star}^{i}$, is comparable
to that of an AGB star, consequently, Eq. {5) prescribes the
condensation-radius to initial-radius -  ratio ($R_{C}^{PCE-RG}/R_{\star}^{i-RG}$)
to be comparable to the ratio of the condensation-radius to star-radius
of an AGB star ($R_{C}^{AGB}/R_{\star}^{AGB}$). Since the radius
at which the bulk mass of the CE mass resides could be comparable
to the radius of an AGB (e.g \citealt{Iva+13,Pass+12,Ric+12}; see Fig.
\ref{fig:AGB-CE comparison}), the CE condensation radius is effectively
located inside its envelope. Moreover, the envelope mass can extend
beyond the condensation radius, and hence be affected by dust-induced
radiation pressure, which can eventually unbind the CE. The evaporation
time is mainly derived from the amount of material which can condense
into dust, the amount of energy radiated on the dust and the amount
of material the dust can push outwards.

We note that convection may also play a role on timescales longer
than the CE inspiral timescale, as in the case explored by \citet{Cla+17},
and may further drive mass from the inner regions outwards beyond
the condensation radius. Indeed, \citet{Cla+17} detailed
evolution model show the existence of non-negligible amount
of envelope mass at sufficiently low temperatures as to potentially allow for
dust condensation. In the next section we will verify this assumption
on the cases of \citet{Cla+17} and \citet{Pass+12}.
 It would be interesting to combine recombination models with the dust-driven models, but this is beyond the scope of the current paper.
\begin{figure*}
\includegraphics[scale=0.1]{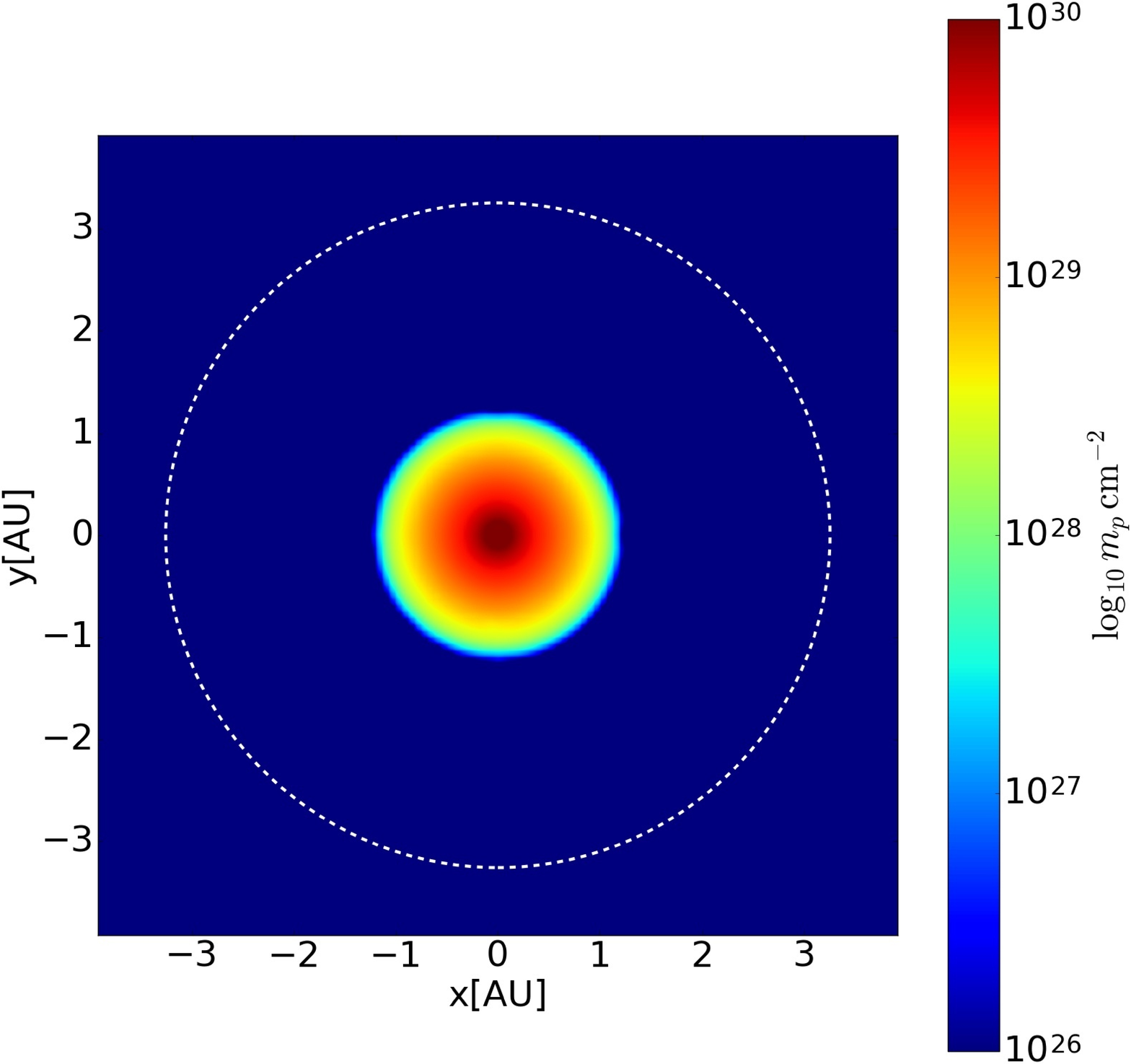}\includegraphics[scale=0.1]{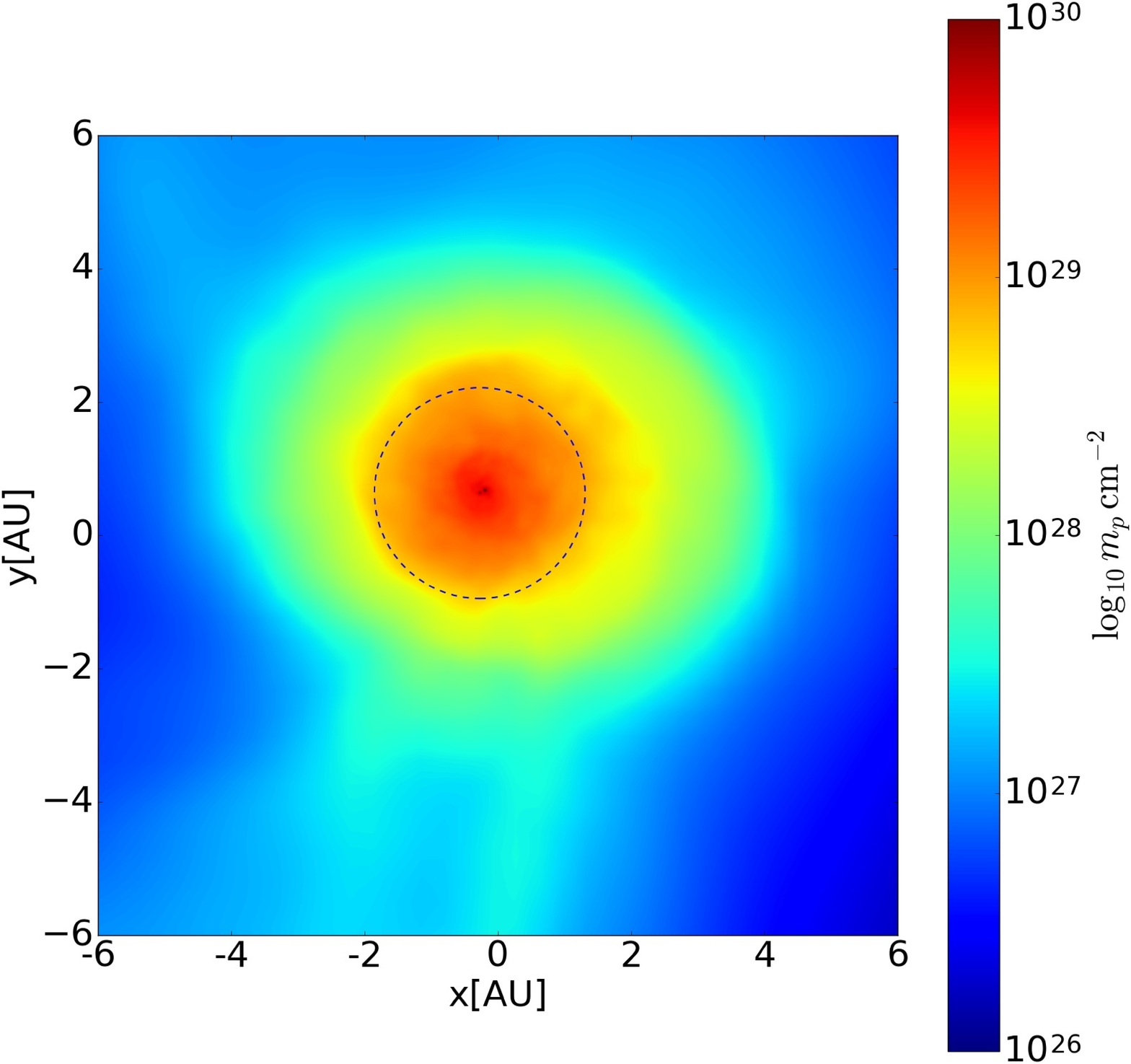}

\includegraphics[scale=0.2]{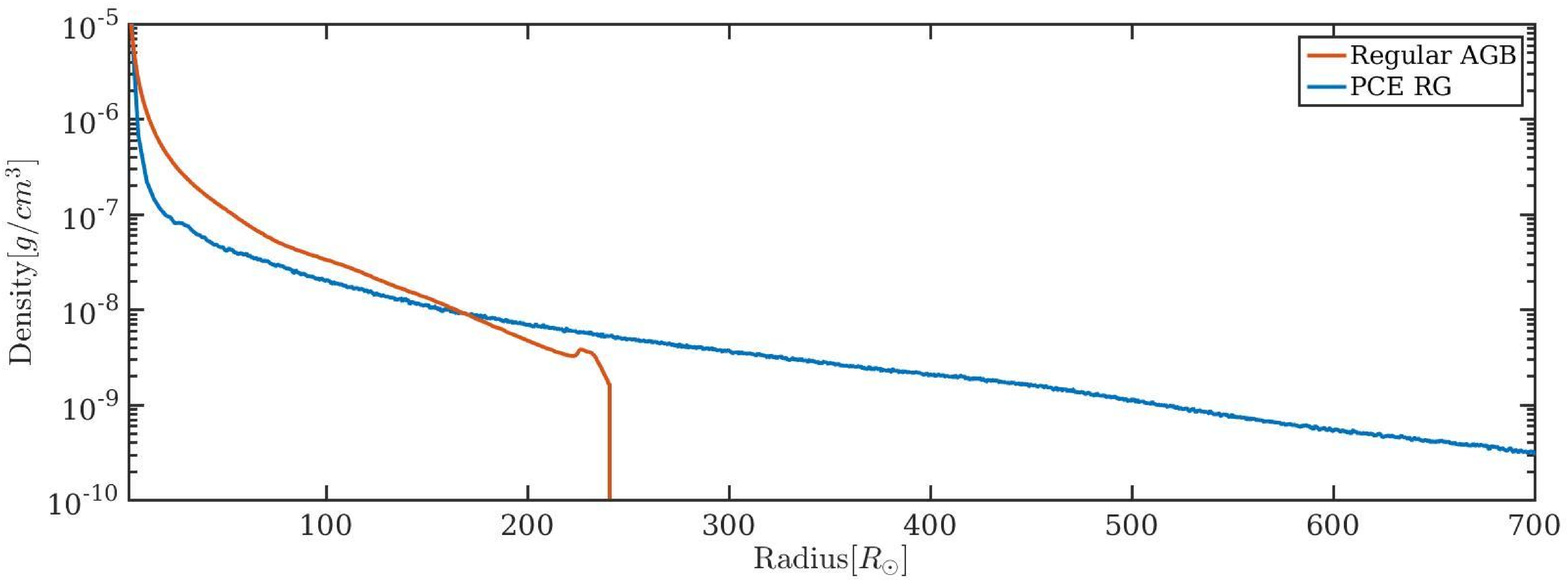}\caption{\label{fig:AGB-CE comparison}A comparison between an evolved AGB
star and a post-CE red giant remnant. Top left: An AGB star evolved
from a Sun-like star of $\sim240\,R_{\odot}$ (using MESA  - \citet{2011ApJS..192....3P}), for which
dust is expected to form at $r\sim700R_{\odot}$ (white dashed
circle), the pulsations in the outwards direction can place material
from the star at that radius, resulting in continuous dust creation
and outwards velocity to the existed grains. Top right: A post Common
Envelope Red Giant star, for which dust formation is expected to form
at $r\sim340R_{\odot}$ (dashed black circle); following the inspiral
of the companion (modeled using Gadget2 - \citet{2005MNRAS.364.1105S}) the gaseous envelope extends
even far beyond the condensation radius. Bottom: Comparison of the
averaged radial density profile of the AGB and post-CE RG stars. As
can be seen, the envelope the post-CE RG extends far beyond the size
of an AGB star, while retaining high densities required for dust formation. }
\end{figure*}

\begin{figure*}
\includegraphics[scale=0.27]{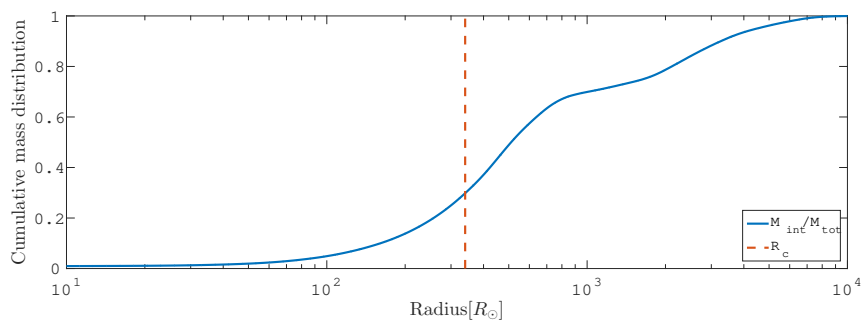}

\caption{\label{fig:Radial Cumulative Mass Ratio} Cumulative radial mass distribution.
The cumulative radial mass distribution of the post-CE RG, showing
that most of the mass of the (still bound) envelope resides beyond
the condensation radius, and is thereby susceptible to dust-assisted
radiation pressure and the production of dust-driven winds. }
\end{figure*}

\subsection{A worked-out example}

\label{subsec:example}

Let us consider a sun-like star which evolved to become a RG with
a radius of $83R_{\odot}$; such a giant has an effective temperature
of $\sim3480K$, luminosity $L\sim1000L_{\odot}$and a hydrogen-exhausted
core of$\sim0.39M_{\odot}$, with a radius of $\sim0.02R_{\odot}$.
Following \citet{Pass+12} we consider the interaction of the RG with
a MS companion leading to a CE phase, resulting in short period stable
orbit of the companion around the RG core, and an extended gas envelope
above it.

In order to obtain the specific parameters of the post-CE RG, and
in particular the density and sound-speed at the condensation radius,
we have repeated the simulation of \citet{Pass+12}, using the MESA
code for following the stellar evolution of the RG until the CE phase,
and using the Gadget2 hydrodynamic code for the hydrodynamical simulation
of the CE evolution. The bloated RG can then be compared with a typical
AGB star. Figure \ref{fig:AGB-CE comparison} compares between the
post-CE RG we obtain and a typical AGB star (as discussed by \citet{Lam+99}
in the context of AGB winds), showing their respective calculated
dust condensation radii. Following the previous sections, the post-CE
RG in this case will have a condensation radius of $R_{C}\approx340R_{\odot}$.
The total mass outside this radius is $M_{env}\sim0.4M_{\odot}$
which consists of almost the entire mass in the CE (see the cumulative
mass distribution in Fig. \ref{fig:Radial Cumulative Mass Ratio}).
A comparison between the density profile the AGB star and the post-CE
RG is presented in the bottom part of Fig. \ref{fig:AGB-CE comparison}.

As described at the previous section, considering the inner part of
the star, below the condensation radius, as a black-body radiation
source, and taking the Rosseland opacity, $\kappa_{Ros}=24(T/T_{C})^{4/3}$,
the temperature in a radius with optical depth $\tau$ is $T\left(\tau\right)^{4}=(3/4)T_{surf}^{4}\left(\tau_{Ros}+\frac{2}{3}\right)$.
The temperature therefore declines from $T=T_{C}=1500K$ at the condensation
radius to $T_{surf}\sim880K$, without consideration of the dust.
The total optical depth, in this case is about $\tau_{tot}\approx10$,
i.e. sufficient for induction of dust driven winds in the comparable
case of an AGB as discussed by \citet{Lam+99}.

The typical time for ejection of the CE in the dust-assisted CE ejection
model is determined by the rate of momentum imparted to the envelope.
In order to eject the envelope one needs to accelerate the envelope
up to escape velocity. Assuming a significant fraction, $f$, of the
momentum exerted by the radiation pressure on the dust is transferred
to the CE, we can compare the accumulated momentum and the total momentum
required to eject the envelope 
\begin{equation}
t_{eject}f\frac{L_{\star}}{c}=M_{env}v_{esc},\label{eq:ejection-time}
\end{equation}
where $t_{eject}$ is the time to eject the envelope, $v_{esc}$ is
the escape velocity, $L_{\star}$is the stellar (core) luminosity
and $c$ is the speed of light. Taking the above values at the condensation
radius we get $t_{eject}\sim1.3\times10^{5}(0.5/f)$ yr; comparable
to the lifetime of an AGB star. Note that for the latter the mass-loss
rate also depends on the rate of pulsations which bring material beyond
the condensation radius. In the CE case this amount depends on the
inspiral phase, where deeper penetration of the companion likely leads
to a larger extension of the CE, and hence more mass expelled beyond
the condensation radius over the inspiral time. Nevertheless, as can
be seen in \citet{Cla+17}, pulsations could be produced
on the longer term evolution, leading to additional expulsion of material
beyond the condensation radius. For an exact calculation of the ejection
time, one should perform an integration over all the values of the
escape velocity ($v_{esc}$) at the region of the envelope above the
condensation radius. 

\section{Discussion and summary}

In this work we have explored a novel scenario for common-envelope
ejection in interacting binaries. We propose that the same dust-driven
winds model suggested for the ejection of AGB stellar envelope could
similarly operate in CE binaries. We show that the temperature and
density conditions in CEs are comparable with those existing in the
envelopes of AGB stars. In particular, most of the CE can extend beyond
the dust condensation radius, following the CE inspiral phase. Hence
dust could efficiently form in the envelope, where it is subjected
to radiation pressure. The gaseous envelope can then be accelerated outwards, forming dust-driven winds that eventually evaporate the CE. 

We considered simple analytic models (similar to those previously
employed for the study of AGB envelopes) to explore dust-driven CE
ejection. More detailed studies using radiative-hydrodynamics modeling,
as had been recently used for the study of AGB envelopes \citep[e.g. ][and references therein]{Fre+17}
could be similarly applied for the study of the CE case. These are
beyond the scope of the current paper which provides a proof of concept
for the dust-driven CE ejection scenario. These will be further explored
in future studies as to provide a more detailed understanding of this
scenario.

We note that we have not discussed the type of the dust grains that
may form in the proposed scenario. Which type of dust grains play
a role in this process depends on the composition of the specific
star. The mechanism of dust driven winds in AGBs works well for C-rich
stars. In O-rich stars significant mass-loss through winds is also
observed, suggesting that a similarly efficient mechanism is at work,
but the dust composition may pose a challenge in this case \citep[see][for an overview]{Hof+15}.
In RGs C-enrichment can only arise from the intrinsic metallicity
of the star, as dredge-up processes occurring in AGB stars do not
take place in this case. However, the same solutions suggested for
O-rich AGB stars such as the role played by micron-sized dust \citep{Hoff+08,Hoff+07}
could similarly operate in this case. Moreover, observations suggest
that dust is efficiently formed even in the merger of main-sequence
stars \citep{Kam+10,Kam+11,Tyl+11,Bar+14,Tyl+15}.

In the dust-assisted CE ejection scenario the timescales for ejecting
the envelope could be long, comparable with the lifetimes of AGB stars.
In practice, CE binaries in this phase may appear very similar to
AGB stars, showing dust enshrouded envelopes, and mass-loss through
slow-winds. Significant non-sphericity of the envelope (e.g. \citealt{Ric+12}), however, could be a distinct signature of post-CEs (possibly observed by \citep{Kam+10,Kam+11,Tyl+11,Bar+14,Tyl+15}. In
the dust-driven CE ejection the CE is not dynamically ejected as typically
envisioned, but rather the dynamical phase introduces the conditions
allowing for the long-term slow mass-loss phase. We point out that
even during the dynamical phase of the inspiral, dust may form and
thereby affect the appearance of the light curve ans spectra of the
transient CEE event and the remnant post-CE binary, and should be
accounted for \citep[see also][for observational evidence and theoretical discussion of the dust effects on the post-CE appearance]{Kam+10,Kam+11,Tyl+11,Bar+14,Tyl+15,Gal+17}
.

Finally, though not the main focus of this paper we note that dust
formation would also assist in trapping radiation arising from recombination
energy, and may further assist any scenario of recombination-energy
induced CE ejection, irrespective of the dust-driven winds explored
here.

\bibliographystyle{mnras}

\bsp	
\label{lastpage}
\end{document}